\documentclass[prb,nopacs,preprintnumbers,twocolumn,amsmath,amssymb]{revtex4}

\usepackage{graphicx}
\usepackage{dcolumn}
\usepackage{color}
\usepackage{bm}
\newcommand{\beq}{\begin{equation}}
\newcommand{\eneq}{\end{equation}}
\newcommand{\bea}{\begin{eqnarray}}
\newcommand{\enea}{\end{eqnarray}}
\newcommand{\bc}{\begin{center}}
\newcommand{\ec}{\end{center}}

\newcommand{\nn}{\nonumber}

\newcommand{\del}{\partial}

\newcommand{\be}{\begin{equation}}
\newcommand{\ee}{\end{equation}}
\newcommand{\eea}{\end{eqnarray}}
\newcommand\beqn{\begin{eqnarray*}}
\newcommand\eeqn{\end{eqnarray*}}
\newcommand{\eqn}[1]{(\ref{#1})}

\begin{document}

\title{Spin connection and boundary states in a topological insulator}
\author{V.Parente$^1$, P.Lucignano$^{1,2}$,  P.Vitale $^{1,3}$, A.Tagliacozzo$^{1,2}$ \\
and \\
F.Guinea$^{4}$}

\affiliation{$^{1}$ Dip. Scienze Fisiche, Universit\`{a} di Napoli Federico II, Via Cintia, I-80126 Napoli, Italy }
\affiliation{$^{2}$  CNR-SPIN, Monte S. Angelo-Via Cintia, I-80126, Napoli,  Italy }
\affiliation{$^{3}$  INFN,  Via Cintia, I-80126, Napoli,  Italy }
\affiliation{$^{4}$  Instituto de Ciencia de Materiales de Madrid (CSIC), Sor Juana In\'es de la Cruz 3, Madrid 28049, Spain }
\begin{abstract}
  We study the surface  resistivity  of a three-dimensional topological insulator when  the  boundaries exhibit a non trivial curvature. We obtain an analytical solution for a spherical topological insulator, and we show that a non trivial quantum spin connection emerges from the three dimensional band structure. We analyze the effect of the spin connection on the scattering by a bump on a flat surface. Quantum  effects induced by the geometry lead to resonances when the electron wavelength is comparable to the size of the bump.
\end{abstract}

\maketitle
\section{Introduction.}
Strong Topological Insulators  (TI 's) are a new class of materials with a bulk gap but surface states defined on surfaces of all orientations\cite{FKM07,FK07,MB07,HK10}, making the boundaries gapless. The number of surface states at a flat surface with a given orientation is odd, and each of them shows a conical singularity, described by the two dimensional Dirac equation\cite{HK10}. Localized states also exist at other lattice defects, such as screw dislocations\cite{RZV09,TK10,TK10b}.

The transport features of electrons at the surfaces of TI 's is being intensively studied. The wavefunctions have an internal spinorial structure made up of two slowly varying components related by time reversal invariance. Backscattering due to smooth perturbations which preserve the time reversal symmetry is forbidden, making the transport properties of these compounds similar to those of graphene in the absence of intervalley scattering\cite{NGPNG09}. Surfaces with a finite curvature allow for scattering processes due to the existence of a non trivial metric, which has been studied in the classical limit\cite{DHAB10}, when wavepackets are well approximated by point particles following classical trajectories.

An analysis of the electronic properties of curved surfaces of TI 's requires information about the way a non trivial metric changes the effective Dirac equation. It is well known that the Dirac equation could be written on a curved space time introducing the spin connection and the rotation of Dirac matrices\cite{N90}. The existence of the spin connection has been postulated in topological insulators\cite{L09}. The emergence of the spin connection from the combination of the three dimensional electronic structure of a TI and the two dimensional metric of a boundary with intrinsic curvature has not been studied so far.

In the next section, we analyze the surface states for the simplest curved boundary with a non trivial metric, the sphere. The conservation of the angular momentum in this geometry allows us to calculate the entire spectrum of surface states, and to show that the spin connection term is induced in the effective surface hamiltonian. We use this information in Section III to analyze the effect of the curvature in the scattering by a bump in a flat surface, a process considered in the classical limit in\cite{DHAB10}. Related processes can be defined in graphene with topological defects\cite{CV07}. Technical details of the calculations are discussed in the appendices, including an analytical study of the boundary states in a cylinder, calculated numerically in\cite{EZL10}. The boundary of a cylinder can be considered a surface without intrinsic curvature and spin connection.

\section{Model of a spherical topological insulator.}
 The surface states of model single particle hamiltonians for a TI have been studied particularly  for a flat boundary \cite{SZS10} and  for an infinite cylinder boundary surface \cite{EZL10}.
A minimal model reproducing the  band structure  of a TI requires four orbitals, related in pairs by the time reversal symmetry\cite{ZI09}. A further simplification includes just  the linear in-momentum contributions to the hamiltonian of Ref. \onlinecite{ZI09}:
\begin{equation}
{\cal H} =\hat{\gamma}^0\Delta+\hbar v_F\hat{\gamma}^ik_i\label{hamil}
\end{equation}
where $v_F$ is the Fermi velocity and the matrices $\hat{\gamma}^a$ are given in terms of Pauli matrices by  $\hat{\gamma}^0=\tau_0 \otimes\tau_z$, $\hat{\gamma}^1=\sigma_x\otimes\tau_x$, $\hat{\gamma}^2=-\sigma_y\otimes\tau_x$  $\hat{\gamma}^3=\sigma_z\otimes\tau_x$. Here $\sigma_a$ and $\tau_b$ denote matrices in the spin and even-odd orbital parity  spaces, respectively. This hamiltonian satisfies time
reversal symmetry $T=\mathcal{K}\: i\sigma_y \otimes\mathbb{I}_{2\times2}$ (here  $\mathbb{I}_{2\times2}$ the ${2\times2}$ identity and $\mathcal{K}$ the complex conjugation). Bulk eigenfunctions in cartesian coordinates are:
\begin{widetext}
\begin{equation}\label{bulk}
\left|\Psi_{1,\pm}\right>=\frac{1}N_{\pm}\begin{pmatrix} \epsilon_\pm ({\vec{\bf{k}}}) +\Delta\\ \hbar v_Fk_z\\0\\ \hbar v_Fk_- \end{pmatrix}e^{i\vec{k}\cdot\vec{r}},\:\:\:\; \left|\Psi_{2,\pm }\right>=\frac{1}N_{\pm}\begin{pmatrix} 0\\ \hbar v_F k_+\\ \epsilon_\pm ({\vec{\bf{k}}})+\Delta\\ -\hbar v_F k_z\end{pmatrix}e^{i\vec{k}\cdot\vec{r}}
\end{equation}
\end{widetext}
 ($k_{\pm}=k_x\pm ik_y$), where  the band energies are $\epsilon_\pm ({\vec{\bf{k}}}) \equiv \pm \sqrt{  \Delta^2 + \hbar ^2 v_F^2 \left(
k_x^2 + k_y^2 + k_z^2 \right)}$ and $N_{\pm}$ is the norm of the states.

Surface states appear in  this model if  the gap parameter, $\Delta$,  changes its sign at the boundary, so that, e.g.,  $\Delta > 0$ in the inside, and $\Delta < 0$ in the vacuum.  The model allows also for the analytical computation of the surface bands of a cylinder\cite{EZL10}, as shown in Appendix~\ref{cylinder}. Later we will include also quadratic corrections to the hamiltonian in Eq.\ref{hamil} and we will show that boundary conditions need to be chosen in a different way in that case.

 In order to obtain the solution of the hamiltonian \ref{hamil} onto a sphere we rephrase its eigenvalue equations into spherical coordinates. The  eigenvector of energy $E$, $\Psi \equiv (\Psi_A,\Psi_B,\Psi_C,\Psi_D)$, satisfies the equations:
\begin{widetext}
\begin{align}
( E - \Delta ) \Psi_A &= i \left[ \cos ( \theta ) \partial_r -
\frac{\sin ( \theta )}{r}
\partial_\theta \right] \Psi_B - i e^{-i\phi} \left[ \sin ( \theta ) \partial_r +
 \frac{\cos ( \theta )}{r} \partial_\theta - i \frac{1}{r \sin (
\theta )} \partial_\phi \right] \Psi_D \nonumber \;,\\
( E + \Delta ) \Psi_B &= i \left[ \cos ( \theta ) \partial_r -
\frac{\sin ( \theta )}{r}
\partial_\theta \right] \Psi_A - i
e^{-i\phi} \left[ \sin ( \theta ) \partial_r  + \frac{\cos ( \theta
)}{r} \partial_\theta - i \frac{1}{r \sin ( \theta )} \partial_\phi
\right] \Psi_C \nonumber \;, \\
( E - \Delta ) \Psi_C &= - i e^{i\phi} \left[ \sin ( \theta )
\partial_r +  \frac{\cos ( \theta )}{r} \partial_\theta + i
\frac{1}{r \sin ( \theta )} \partial_\phi \right] \Psi_B - i \left[
\cos ( \theta ) \partial_r -  \frac{\sin ( \theta )}{r}
\partial_\theta \right] \Psi_D \nonumber \;, \\
( E + \Delta ) \Psi_D &= - i e^{i\phi} \left[ \sin ( \theta )
\partial_r +  \frac{\cos ( \theta )}{r} \partial_\theta + i
\frac{1}{r \sin ( \theta )} \partial_\phi \right] \Psi_A - i \left[
 \cos ( \theta ) \partial_r -  \frac{\sin ( \theta )}{r}
\partial_\theta \right] \Psi_C\:,
\label{hamil_r}
\end{align}
\end{widetext}
(here  $\hbar=v_F = 1$), with  the boundary conditions
\begin{align}
\Delta ( r , \theta , \phi ) = \left\{ \begin{array}{lr} \Delta _{in}&r<R \\ \Delta _{out}&r>R \end{array} \right.
\end{align}
We choose $  \Delta _{in} =-\Delta _{out}= \Delta $ for simplicity, so that  the exponential decay of the boundary states into the bulk near a flat surface is defined by the length scale $\Lambda = \hbar v_F/ \Delta$. The angular momentum is conserved and  is quantized in half integer units (see, for instance Ref. \onlinecite{GGV92}). Its eigenfunctions allow us to reduce the set of Eq.s(\ref{hamil_r}) to two coupled differential equations for the radial coordinates, as discussed in Appendix B.
\begin{figure}
\begin{center}
\includegraphics[width=6cm]{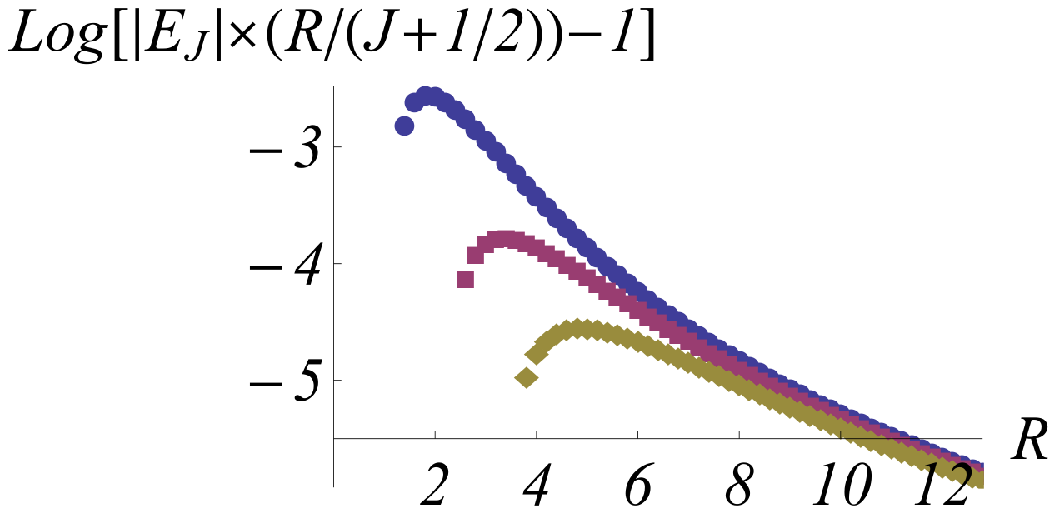}
\includegraphics[width=6cm]{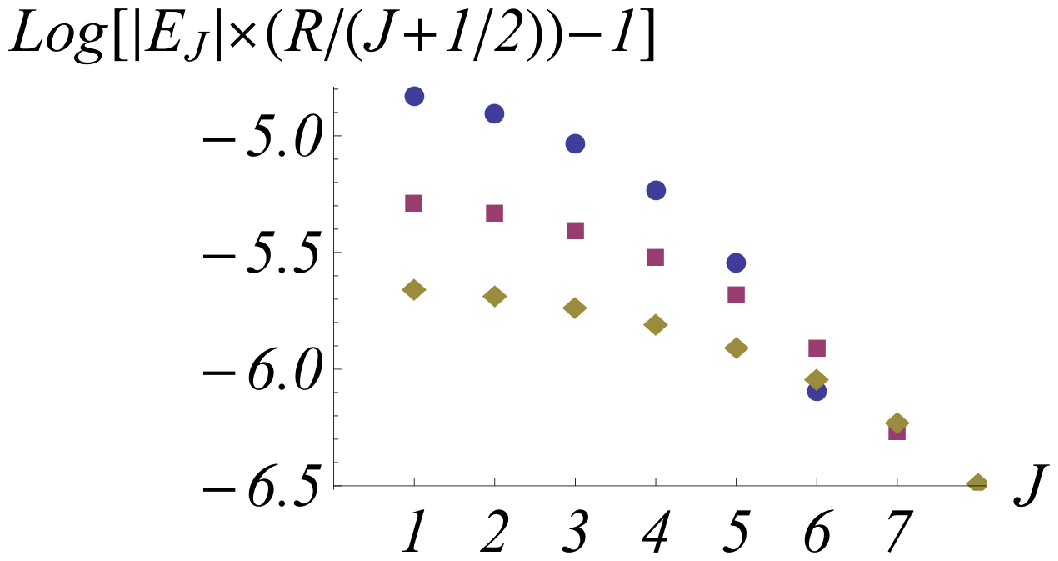}
\caption{(Color online). Dependence of the surface energy levels on angular momentum, $J$, and radius, $R$. The deviation from the result for the two dimensional Dirac equation on a sphere, see eq.~\ref{levels} is plotted. Top: Dependence on $R$. From top to bottom, $J=1,2,3$. Bottom: Dependence on $J$. From top to bottom, $R = 8, 10, 12$. In all cases, $v_F=1$ and $\Delta = 1$ .}
\label{scaled_levels}\end{center}
\end{figure}
It can be shown  that the energy spectrum converges exponentially to the one of  the two-dimensional Dirac equation onto a sphere:
\begin{align}
E_J &= \pm \frac{\hbar  v_F ( J + 1/2 )}{R} \times \left[ 1 + {\cal{O}} \left ( e^{-  R/\Lambda } \right )  \right] \nonumber \;,\\  J &= \frac{1}{2} , \frac{3}{2} \cdots J_{max} \;.
\label{levels}
\end{align}
where $J_{max} \sim  R/\Lambda $. The multiplicity of each level is $2 J + 1$. The exponential convergence of the energy levels to the asymptotic value in eq.~\ref{levels} is shown in Fig.~\ref{scaled_levels}. This type of convergence implies that the effective hamiltonian describing the surface modes does not admit an expansion on higher order derivatives, of the type $\Delta ( \Lambda \partial_i  )^n$. The study of the hamiltonian in eq.~\ref{hamil} can be extended in a straightforward way to the case when $\Delta_{out} \ne \Delta_{in}$, although it becomes cumbersome to obtain analytical expansions in the limit $R \rightarrow \infty$. The numerical solution, obtained by generalizing the analysis of Appendix B, shows an agreement with the spectrum in  Eq.~\ref{levels} of the same accuracy as those reported in Fig.~\ref{scaled_levels}.

\begin{figure}
\begin{center}
\includegraphics[width=6cm]{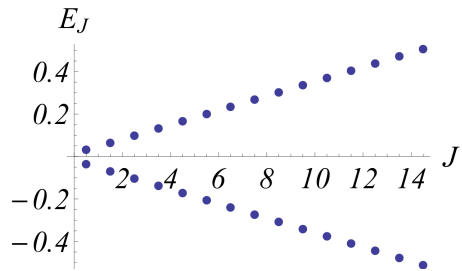}
\caption{(Color online). Energy levels of a spherical topological insulator of radius $R=30$ with a quadratic dispersion relation, obtained by the replacement $\Delta \rightarrow \Delta + \alpha ( k_x^2 + k_y^2 + k_z^2 )$ in eq.~\ref{hamil}. Other parameters are $\Delta=1$ and $v_F = 1$. The boundary conditions are $\Psi_A ( R ) = \Psi_B ( R ) = \Psi_C ( R ) = \Psi_D ( R ) = 0$.}
\label{levels_alpha}\end{center}
\end{figure}

As quadratic terms  do not break the spherical symmetry, they can be safely  added to Eq.~\ref{hamil}, by  the simple  substitution $\Delta \rightarrow \Delta + \alpha  ( k_x^2 + k_y^2 + k_z^2 )$, where $\alpha$ is a constant. Hence the angular part of the wavefunctions remains unchanged while its radial part satisfies  second order coupled equations, in place of those in Eq.~\ref{hamil_r}. For each value of the energy, $E$, we find evanescent waves with two different decay lengths, $ \Lambda _1 ( E )$ and $\Lambda _2 ( E )$, which are given by the roots of a fourth order polynomial. The boundary conditions need to be replaced. The simplest boundary condition compatible with the new second order equations is $\Psi_A ( R ) = \Psi_B ( R ) = \Psi_C ( R ) = \Psi_D ( R ) = 0$\cite{SZS10}. By solving numerically these boundary conditions, we find again an agreement with eq.~\ref{levels} similar to that shown in Fig.~\ref{scaled_levels}. Results are shown in Fig.~\ref{levels_alpha}.

We conclude that the boundary states on a spherical TI  satisfy the Dirac equation on the surface of the sphere. The spin connection, related to the intrinsic curvature of the metric, clearly emerges at the boundaries of a TI.  More generally, the boundary states satisfy the  Dirac equation on a curved space-time \cite{N90}
\begin{equation}
\gamma^{\mu}(\partial_\mu+\Gamma_\mu)\Psi=0 \:\: .
\label{curvata}
\end{equation}
Here  $\gamma^\mu=e_a^{\;\;\;\mu}\gamma^a $ ( with $\gamma^{a=0} = -i\sigma ^z $,$\gamma^{a=1} = \sigma ^y $ and $\gamma^{a=2} = - \sigma ^x $ )  are the rotated Dirac matrices, satisfying the  generalization of the flat algebra
\beq \label{algcur}
\left \{ \gamma ^\mu , \gamma ^\nu \right \} = 2 g ^{\mu \nu} \:\: .
\eneq
 The Minkowski metric $ \eta _{ab} $ has been  replaced by the curved one $g_{\mu\nu} = \eta _{ab} \: e^a_{\;\;\mu}e^b_{\;\;\nu} $. The tetrads $e^a_{\;\;\;\mu}$ are the elements of the Jacobian matrix of  the transformation from the coordinates $x^\mu$, defined on the whole manifold, to a local inertial frame:
\begin{equation}
e^a_{\;\;\;\mu}=\frac{\partial x^a}{\partial x^\mu} \:\: .
\end{equation}
 $\Gamma_\mu$ is the spin connection  $\frac{i}2\Gamma^{a\;\;\;b}_{\;\;\;\mu}\Sigma_{ab}$ where   $\Sigma_{ab}=i/2[\gamma_a,\gamma_b]$ are the generators of the spinorial representation of the Lorentz group, and the connection coefficients   $\Gamma^{a\;\;\;b}_{\;\;\;\mu}$ are given by
$\Gamma^{a\;\;\;b}_{\;\;\;\mu}=e^a_{\;\;\;\nu}\nabla_\mu e^{b\nu}$.

 \section{Scattering off a gaussian bump}
 \subsection{Unrelaxed lattice}
We now derive the resistivity for electrons propagating  at the flat boundary surface of a TI, when they are scattered off  a gaussian bump of  height $z\left (|\vec{r}|\right ) = h \: e^{- r^2/\ell^2 } $.
The Boltzmann relaxation-time approximation can be used ($\nu  ( 0 ) = k_F /(\pi \hbar v_F ) $ is  the density of states at the Fermi level for both spins):
\beq
\rho \left (k_F \right ) = \frac{2}{e^2 v_F^2 \nu ( 0 ) } \: \frac{ 1}{\tau ({k}_F ) } \:\: ,
\label{resist}
\eneq
 where  the  usual  definition of the total relaxation rate is
\begin{eqnarray}
\frac{1}{\tau (k_F) }
=\frac{2\pi }{\hbar }   \nu (0) \int_0^{2\pi}  d\theta \left ( 1- \hat{k}\cdot \hat{k}' \right )  \left| \left<k| t^{eff} |k' \right>\right |^2. \:\: \hspace{0.5cm}
\label{relt}
\end{eqnarray}
Here $\left<k| t^{eff} |k' \right>$ is the matrix element of the $t-$matrix, which depends on the energy and on the scattering angle $ \theta$ between the incoming and outgoing wave.

Since the metric  induced on the manifold  by the bump is axially symmetric, it is convenient to rewrite  the two dimensional Dirac equation in flat space time
 in cylindrical coordinates:
 \beq
-i \hbar v_F \left ( \sigma ^r \partial _r + \sigma
^\theta \:  \frac{1}{r} \partial _\theta  \right ) \Psi=E\Psi \:\: .
\label{flat}
\eneq
Here the matrices $\sigma ^{r,\theta } $ are $
\sigma ^r = \cos \theta  \:\sigma ^x + \sin \theta \: \sigma ^y $ and
$\sigma ^\theta = -\sin \theta  \:\sigma ^x + \cos \theta \: \sigma ^y $.

Given the metric\cite{JCV07}
 \bea
 g_{\mu\nu} =  \left ( \begin{array} {c c c} -1 & 0 & 0  \\ 0 & 1 + f(r) &  0 \\ 0 & 0  & r^2  \end{array} \right )   \:\: ,
 \label{metric}
  \enea
where   $f(r)=\left (dz(r)/dr\right )^2$, we rewrite Eq.\eqref{curvata} and, as shown in Appendix C,  the spin connection  $\Gamma_\mu$ can be embodied  in  the wavefunction as a real  prefactor $ \Psi=\Phi  \:  \exp { \int_r^{+\infty} dr'\: A_\theta (r' ) } $,  where $A_\theta(r)=\frac{1}{2 r}\left( \sqrt{1+f(r)}- 1\right)$. The spinor $\Phi$ satisfies the equation
\begin{equation}
-i \left[\frac{\sigma^r}{\sqrt{1+f(r)}}\frac{\partial}{\partial r}+\sigma^\theta\frac{1}{r}\frac{\partial}{\partial\theta}\right]
\Phi= s k  \: \Phi \:\: .
\label{scur}
\end{equation}
\\
where $s= + (-) $ for particles (holes)  and $E= \hbar v_F k $. The real prefactor can be interpreted as the origin of
 charge puddles  accumulating at the bump.
 Eq. \eqref{scur} describes an unrelaxed lattice.  Relaxation of the structure, besides adding an effective gauge potential, may further change the  spin connection.   As  elastic deformations do not add any curvature, the  change only implies  a trivial holonomy on the wave function.  This is a way of restating the Saint Venant conditions for the two-dimensional case.
Changes in the Dirac Equations  are well localized in space close to the bump, hence a scattering picture can be fruitfully adopted here. We will focus on  the particle sector of the theory,  and assume  that the incoming $k-$vector is in the direction of the polar  axis ($\theta  = 0 $). The eigenfunctions can be expressed as superposition of  angular momentum $m$ eigenstates:
\begin{equation}\label{eq:ansatz}
\Phi_m(r,\theta)=
\left ( \begin{array}{c}
u_m(r)\\
iv_m(r)e^{i\theta}
\end{array} \right )
e^{im\theta}  \:\: .
\end{equation}
The Born approximation  is worked out, to lowest order, in Appendix  C. Using an asymptotic expansion of the wavefunctions given in Eq.\eqref{born1}, we have:
\begin{widetext}
\begin{align}
  \frac{ 1}{\tau \left (\vec{k}_F \right ) } &= \frac{n_b v_F}{k_F} \times \sum_{\substack{m}}  \:\left [ \sin ^2  \delta _m   -  \cos { \left (\delta _{m+1} - \delta _{m-1 }  \right )\sin\delta_{m+1}\sin\delta_{m+1}}   \:  \right ]
  \label{reltB}
\end{align}
\end{widetext}
where the phase shifts $\delta_m$ for the $m^{th}$ component of the wavefunction are reported in appendix C and $n_b$ is the concentration of bumps.

At low incoming electron energy,  it turns out that   the terms with  $m= 0,\pm 1, \pm 2 $ are
 ${\cal{O}}[(k\ell )^4 ] $ and when  choosing  $4\pi ^2 h^2/\ell ^2 \sim 1 $, they  sum up to ${\cal{S}}\approx 0.733 $.
 The terms with  $m= \pm 3 $ are   ${\cal{O}}[(k\ell )^8 ] $, while  the terms  $m= \pm 4 $  are  ${\cal{O}}[(k\ell )^{12} ] $ (see Appendix C).

\begin{figure}[htbp]
   \centering
    \includegraphics[width=0.9 \linewidth]{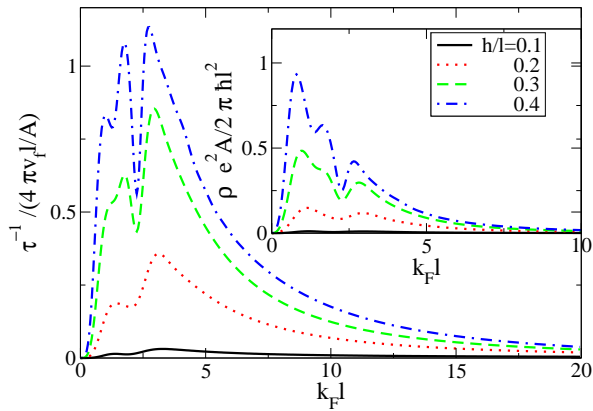}
   \caption{(Color on-line) Main Panel: inverse scattering time as a function of $k_Fl$ for different values of the aspect ratio of the bump $h/l$. Inset: the resistivity due to scattering off a bump on the surface of a topological insulator, in units  $ \rho  \cdot e^2 A / ( 2\pi \hbar \ell ^2 )  $ vs. dimensionless energy $k\ell$, at different ratios $h / \ell$. }
   \label{fig}
\end{figure}

 Eventually, the resistivity for independent  point like  defects, when the carrier density is low, (i.e. low incoming energy) is:
  \beq
\rho \left (k_F \right )   \sim   \frac{ 2h }{e^2  }\:  n_b \pi  \ell ^2 \: \left \{   \: {\cal{S}}\left (\frac{h}{\ell} \right ) \: (k_F\ell)^2   + {\cal{O}}\left [(k_F\ell )^4\right  ] \right \} \: .
\label{rebu}
\eneq
$ {\cal{S}}\left ( h/\ell  \right )$ is a numerical prefactor which depends on the   strength of the perturbation parametrized by $ h/\ell$. The plot  of the resistivity   $vs$ energy in dimensionless units, $k_F \ell $, for various values of  $ h/\ell$  is shown in Fig. \ref{fig}. The leading term  is proportional to the density of carriers $n$.

It has been proven recently that the classical limit for large incoming energy  (i.e. relatively high densities $n$) corresponds to an energy independent  $v_F\: \tau (k_F) $ \cite{DHAB10}. This implies that, according to Eq.\eqref{resist},  $ \rho \propto h/(e^2 k_F) \sim 1/\sqrt{n} $, in this limit.  We derive the same conclusion with a careful analysis of  the sum in Eq.\eqref{reltB} at $k\ell >>1 $.  Classically, angular momentum conservation in the scattering implies that  $ m  \sim k  b$. Here $b$ is the impact parameter measured from the center of the bump in the direction orthogonal to $\vec{k}$. By asymptotically expanding the Bessel functions appearing in the phase shifts,  it turns out that there is a collection of terms contributing to the sum, which are roughly independent of $m$, as long as $k\ell > >  m $.  For these $m$ values,  $\tan \delta _m $ is   of the form:
  \bea
\tan \delta _m \approx   \pi \: \left (\frac{h}{\ell}\right ) ^2\: \frac{  k\ell }{2} \:  \left [    \sqrt{\frac{\pi}{2}} - (-1)^m   \frac{2}{(k\ell )^3 } \right ]  \: .
   \enea
   All other terms  scale  for $m >> k\ell $  as
\beq
  \sqrt{\frac{m}{2}}  \: \left (\frac{k\ell}{2}\right)^{2(m+1)} \frac{1}{(2m-1)!!}
 \eneq
and therefore they  rapidly converge to zero.

We conclude that in the semiclassical limit   $k\ell \gg  1$  a factor $k_F $ comes from the relevant terms in the sum of Eq.\eqref{reltB}, which are all of the same order. This factor cancels with the $k_F$ appearing in the denominator, so that the result  for $v_F \tau (k_F) $ is independent of $k_F$.  This is in fact found numerically. In Fig. \ref{fig} we see that  the conductivity  $ \sim \rho ^{-1}$  grows linearly at large $k_F\ell$ and has a minimum in the neighborhood of $k_F\ell \sim 1$.  In Appendix D we report a simple argument based on a saddle point approximation of  the $\sum _m $ which qualitatively recovers the classical limiting result for large $k_F\ell$, derived in Ref.~\onlinecite{DHAB10}. In Fig. \ref{fig} it is shown a significant increase of the cross section of the bump for $k_F \ell \sim 1$ and $h / \ell \gtrsim 0.2$. This increase is due to quantum resonances induced by the non trivial spin connection.

\section{Discussion.}
We have shown in Section II how the two dimensional Dirac equation in curved space emerges at the simplest boundary with non trivial metric, the surface of a sphere. The metric enters through the spin connection, which reflects the properties of the internal spin under parallel transport along the surface\cite{N90}. The spin connection reflects a quantum feature of the electrons, and cannot be inferred from solely classical arguments. The spin connection leads to a finite Berry phase when the electron is transported around a closed geodesic. A manifestation of this effect is the quantization of the total angular momentum in half integer units.

The model that we have studied leads to simple analytical expressions of the energies and wavefunctions of the boundary states. They can be used as a zeroth approximation to situations close to spherical symmetry, or where an isotropic electronic structure can be obtained by rescaling a length. We find that the corrections to the two dimensional Dirac equation depend exponentially on $ R/\Lambda \equiv R\Delta  /\hbar  v_F$\cite{fullerenes}.

The model describes external surfaces of mesoscopic crystals and internal voids in bulk systems.  In the case of a small void, we find that two doublets at a finite distance of the Dirac energy appear for radii $R \gtrsim 2 /\Lambda $. These voids will act as molecules embedded into the bulk material. The interaction energy of electrons localized inside the voids scales as the level separation, $E_{int} \approx e^2 / ( \epsilon R )$, where $\epsilon$ is the dielectric constant of the topological insulator. At temperatures below this scale, voids with an odd number of electrons will give rise to magnetic moments. The RKKY interaction between moments at different vacancies should decay exponentially, $J_{RKKY} ( \vec{\bf r} - \vec{\bf r}' ) \sim e^{-  | \vec{\bf r} - \vec{\bf r}' |  / \Lambda }$. Hence, small vacancies might give rise to a paramagnetic susceptibility in topological insulators. If they are within a distance $d \sim \Lambda $ from the surface, these local moments will hybridize with the surface states, leading to the Kondo effect\cite{FCGWZ10,Z10}.

We have analyzed the scattering of Dirac fermions by surface corrugations which induce a non trivial curvature in the quantum limit, $k_F \ell \lesssim 1$. We find that the resistivity due to a finite concentration of bumps, $n_b$, vanishes as $k_F^2$ for small $k_F$, due to a combination of a density of states factor, which goes as $k_F$, and a scattering time which increases as $k_F^{-3}$. By comparison, the scattering time in the classical regime\cite{DHAB10} ($k_F \ell \rightarrow \infty$) is independent of $k_F$, and $\rho \sim k_F^{-1}$. The wave nature of the quasiparticles allow them to diffract around the bump, making it effectively transparent for long wavelengths, $k_F \ell \ll 1$. The non trivial curvature induces quantum reseonances for $k_F \ell \sim 1$ and an aspect ratio $h / \ell \gtrsim 0.2$.
\begin{acknowledgments}
We acknowledge important discussions with  A. Akhmerov, R. Egger and M. A. H. Vozmediano.  We acknowledge financial support from MIDAS (Macroscopic Interference Devices for Atomic and Sold State Physics) and from MAMA (Multifunctioned Advanced Materials and nanoscal phenomena) F.G. is supported by
MICINN (Spain), Grants FIS2008-00124 and CONSOLIDER CSD2007-00010.
\end{acknowledgments}
\appendix
\section{Boundary states at cylinder surface}
\label{cylinder}
In the following appendixes we will use $\hbar=v_F=1$, except in main results. Let us start from the $\vec{k}\cdot \vec{p} $   model   Hamiltonian of Eq.\eqref{hamil}:
\begin{equation}
H \left [ \vec{r} \right ] =
\begin{pmatrix}
\Delta &i\partial_z & 0 &i(\partial_x +i\partial_y)\\
i\partial_z & -\Delta & i(\partial_x +i\partial_y)& 0\\
 0 &i(\partial_x -i\partial_y)&\Delta & -i\partial_z \\
 i(\partial_x -i\partial_y) & 0& -i\partial_z  & -\Delta
\end{pmatrix}.
\label{hamcyl}
\end{equation}
To find surface states in this approximation is enough to match the solutions of the Schr\"odinger equation at the surface of the cylinder. The gap $\Delta $   should  change its sign   between in and out of the surface.
We rewrite the eigenvalue problem  in cylindrical coordinates for the 4-component spinor $(\Psi_A ,\Psi_B,\Psi_C,\Psi_D )$:
\begin{widetext}
\begin{align}
( E - \Delta )\: \Psi_A ( \vec{\bf r} ) &=  i
\frac{\partial}{\partial_z} \Psi_B ( \vec{\bf r} ) + e^{i\theta }   \left( i
\frac{\partial}{\partial r} - \frac{1}{r}
\frac{\partial}{\partial \theta} \right)
\Psi_D ( \vec{\bf r} ) \nonumber \\
( E + \Delta ) \:\Psi_B ( \vec{\bf r} ) &= i
\frac{\partial}{\partial_z} \Psi_A ( \vec{\bf r} )   + e^{i\theta }  \left( i
\frac{\partial}{\partial r} - \frac{1}{r}
\frac{\partial}{\partial\theta} \right) \Psi_C ( \vec{\bf r} )
\nonumber
\\ ( E - \Delta ) \: \Psi_C ( \vec{\bf r} ) &=  e^{-i\theta }  \left( i
\frac{\partial}{\partial r} + \frac{1}{r}
\frac{\partial}{\partial \theta } \right) \Psi_B ( \vec{\bf r} )  - i
 \frac{\partial}{\partial_z} \Psi_D ( \vec{\bf r} ) \nonumber
\\ ( E + \Delta )  \: \Psi_D ( \vec{\bf r} ) &=  e^{-i\theta }  \left( i
\frac{\partial}{\partial r} + \frac{1}{r}
\frac{\partial}{\partial \theta } \right) \Psi_A ( \vec{\bf r} ) -i
 \frac{\partial}{\partial_z} \Psi_C ( \vec{\bf r} )
\label{dirac_cylinder}
\end{align}
 (in the following  $k$ is in the $\hat z$ direction, which is the axis of the infinite cylinder.
Inside the cylinder, the wavefunctions which are mostly localized close to  the surface involve the modified   Bessel functions  $I_n ( \kappa r )$ with integer $n$. They  diverge
exponentially at infinity   but are finite for  $r \rightarrow 0$. The two eigenfunctions at fixed energy $E$ are ($\kappa$ is  unknown for the moment):
\begin{equation}
\left| E, 1_<\right>=\frac{1}N\begin{pmatrix}( E+\Delta )\:  I_{n} ( \kappa r )\\ k \:  I_{n} ( \kappa  r )\\0\\i\kappa  \: I_{n+1} ( \kappa r )  \: e^{-i \theta}\end{pmatrix} \:  e^{-i n\theta}\:  e^{-ik z},
\; \left|E, 2_<\right>=\frac{1}N\begin{pmatrix} i\kappa \:  I_{n} ( \kappa  r ) e^{i \theta} \\0\\-k \: I_{n+1} ( \kappa r ) \\ (E-\Delta)\: I_{n+1} ( \kappa r )  \end{pmatrix}
 \:  e^{-i (n+1)\theta}\:  e^{-ik z}.
 \label{surcyli}
\end{equation}
  The energies of these states are
$E = \pm \sqrt{\Delta^2 +  k^2 -\kappa^2} $.
Outside  the cylinder,  the functions $K_n ( \kappa r)$  replace the $I_n(\kappa r)$,  as the former  decay exponentially  for $\kappa r \rightarrow \infty$ and  $\Delta \to -\Delta $.
 The eigenfunctions are:
\begin{equation}
\left| E, 1_>\right>=\frac{1}N\begin{pmatrix}( E-\Delta )\:  K_{n} ( \kappa r )\\ k \:  K_{n} ( \kappa  r )\\0\\-i\kappa  \: K_{n+1} ( \kappa r )  \: e^{-i \theta}\end{pmatrix} \:  e^{-i n \theta}\:  e^{-ik z},
\; \left|E, 2_>\right>=\frac{1}N\begin{pmatrix} - i\kappa \:  K_{n} ( \kappa  r )  \: e^{i \theta}\\0\\ -k\: K_{n+1} ( \kappa r )  \\
(E+\Delta) \: K_{n+1} ( \kappa r )\end{pmatrix}
 \:  e^{-i (n+1)\theta}\:  e^{-ik z}.
 \label{surcylk}
\end{equation}
\end{widetext}
The eigenvalues are again those of Eq.\eqref{surcyli}

The two wavefunctions inside the cylinder should be matched  to the two outside
for each value of $n$. The matching conditions  at $R$,  the radius of the cylinder,  lead to
\begin{widetext}
\begin{align}
{\rm Det} \left| \begin{array}{cccc} i  \kappa I_n ( \kappa R )
&( E + \Delta ) I_n ( \kappa R ) &- i  \kappa K_n ( \kappa R ) &(
E - \Delta ) K_n ( \kappa R ) \\ 0 & k I_n ( \kappa R ) &0 & k
K_n ( \kappa R ) \\ -  k I_{n+1} ( \kappa R ) &0 &-  k K_{n+1}
( \kappa R ) &0 \\ ( E - \Delta ) I_{n+1} ( \kappa R ) &i  \kappa
I_{n+1} ( \kappa R ) & ( E + \Delta ) K_{n+1} ( \kappa R ) &-i
\kappa K_{n+1} ( \kappa R ) \end{array} \right| &= 0.
\label{det_cylinder}
\end{align}
The vanishing of the determinant implies:
\beq
\left [ I_{n}^2 \:  K_{n+1}^2 + K_{n}^2 \:  I_{n+1}^2 \right ] \: \kappa ^2 + \left ( 2\kappa ^2 - 4 \Delta ^2 \right ) \:
I_{n} \:  I_{n+1} K_{n} \:  K_{n+1}  = 0 \: .
\label{deto}
\eneq
\end{widetext}
The presence of products of  the Bessel functions $I_{n}$ and $K_{n}$ assures that, in the limit of
 $ \kappa R >>1 $, there is just an inverse powerlaw dependence  of the secular problem on $ \kappa R$.
 and  Eq.\eqref{deto} becomes:
\beq
\kappa ^2  = \Delta ^2  \: \left [ 1-\: \frac{(n+1/2)^2}{\Delta^2 R^2}\right ] \:\: .
 \eneq
Hence,   the energy of the states is, according to Eq.\eqref{eig}
\begin{align}
E &= \pm \hbar v_F \: \sqrt{k^2  +\frac{(n+1/2)^2}{ R^2}  }  + {\cal{O}} \left ( \frac{\hbar v_F^2}{\Delta ^2R^2} \right ) &R \gg \frac{\hbar v_F}{\Delta}
\label{eig}
\end{align}
 This result is in complete agreement with Eq.5 of Ref. \onlinecite{EZL10}.
 Let us now consider the opposite limit $\kappa R << 1 $. Expansion gives, up to second order in $1/\kappa R $:
\beq
 \kappa^2 \:\left \{ 2n(n+1) \left [ 1+ \frac{1}{16 \: n^2 (n+1)^2 } \right ] + \frac{1}{2} \right \}  = \Delta ^2 \:\: .
 \eneq
 The energy reads in this limit:
 \begin{widetext}
 \beq
E_n ( k ) \approx   \pm\Delta  \sqrt{  1-  1   / \left ( 2n(n+1) \left [ 1+ \frac{1}{16 \: n^2 (n+1)^2 } \right ] + \frac{1}{2} \right )  +\left  (\frac{\hbar v_F k_z }{\Delta}\right  )^2 } \:\:
 + {\cal{O}} \left ( \frac{\hbar \Delta R}{v_F} \right )  \:\:\:\: for \:\:\: R \ll \hbar v_F / \Delta \:\: .
\eneq
\end{widetext}
\section{Angular momentum eigenstates}
\label{angular}
Generalized   angular momentum  operators $J$ can be defined as  usual as the sum of spin and orbital angular momentum.
It can be shown that the Hamiltonian in eq.~\ref{hamil_r} commutes with $J^2,J_z$ therefore its eigenstates can be labeled by  $| j , m \rangle$ with  $\vec{\bf J}^2 | j , m \rangle = j ( j + 1) | J , J_z \rangle$ and $J_z | j , m \rangle = m | J , J_z \rangle$. In order to obtain single valued eigenfunctions, the values of $J$ and $J_z$ must be half integers.
As usual by using $J^+ | j ; j \rangle = 0$ and $J^- | j , m \rangle \propto | j , m - 1 \rangle$ we can explicitly construct the wavefunctions of the different states $| j , m \rangle$. The Hamiltonian eigenfunction can be thus expanded onto the the lowest angular momenta states:
\begin{widetext}
\begin{align}
\left| \frac{1}{2} \, \, , \, \, \frac{1}{2} \right\rangle &= A
\left(
\begin{array}{c} -  \cos ( \theta ) \\0 \\ \sin ( \theta ) e^{i \phi}
\\ 0 \end{array} \right) +
B \left(
\begin{array}{c} 0 \\ -  \cos ( \theta ) \\0 \\ \sin ( \theta ) e^{i \phi}
 \end{array} \right) + C \left( \begin{array}{c} 1 \\ 0 \\ 0 \\ 0 \end{array} \right)
 + D \left( \begin{array}{c} 0 \\ 1 \\ 0 \\ 0 \end{array} \right) \nonumber  \\
\left| \frac{1}{2} \, \, , \, \, - \frac{1}{2} \right\rangle &= A
\left(
\begin{array}{c}   \sin ( \theta ) e^{-i \phi} \\0 \\  \cos ( \phi
) \\ 0 \end{array} \right) + B \left(
\begin{array}{c} 0 \\ \sin ( \theta ) e^{-i \phi} \\0 \\  \cos ( \phi )
 \end{array} \right) + C \left( \begin{array}{c} 0 \\ 0 \\ 1 \\ 0 \end{array} \right)
 + D \left( \begin{array}{c} 0 \\ 0 \\ 0 \\ 1 \end{array} \right) \nonumber  \\
 \left| \frac{3}{2} \, \, , \, \, \frac{3}{2} \right\rangle &= A
\left( \begin{array}{c} -  \sin ( \theta ) \cos ( \theta ) e^{i
\phi} \\ 0 \\ \sin^2 ( \theta ) e^{2 i \phi} \\ 0 \end{array}
\right) + B \left( \begin{array}{c} 0 \\ -  \sin ( \theta ) \cos (
\theta ) e^{i \phi} \\ 0 \\ \sin^2 ( \theta ) e^{2 i \phi}
\end{array} \right) + C \left( \begin{array}{c} \sin ( \theta ) e^{i \phi} \\ 0 \\ 0 \\ 0 \end{array}
\right) + D \left( \begin{array}{c} 0 \\ \sin ( \theta ) e^{i \phi} \\ 0 \\ 0 \end{array} \right) \nonumber \\
\left| \frac{3}{2} \, \, , \, \, \frac{1}{2} \right\rangle &= A
\left( \begin{array}{c} - 2 \cos^2 ( \theta ) + \sin^2 ( \theta ) \\
0
\\ 3  \sin ( \theta ) \cos ( \theta ) e^{i \phi} \\ 0 \end{array}
\right) + B \left( \begin{array}{c} 0 \\ - 2 \cos^2 ( \theta ) +
\sin^2 ( \theta ) \\ 0
\\ 3  \sin ( \theta ) \cos ( \theta ) e^{i \phi} \end{array}
\right) +
 C \left( \begin{array}{c} 2 \cos ( \theta
) \\ 0 \\ \sin ( \theta ) e^{i \phi} \\ 0 \end{array} \right) + D
\left(
\begin{array}{c} 0 \\ 2 \cos ( \theta ) \\ 0 \\ \sin ( \theta ) e^{i
\phi} \end{array} \right)
 \end{align}
\end{widetext}
where the states $|3/2,-1/2\rangle,|3/2,-3/2\rangle$  are not explicitly exhibited, here.
\section{Spherical boundary states and energy spectrum}
It can be snown that boundary states in the spherical case for $j=m = n-1/2$  ($n >0 $) have the following form:
\begin{widetext}
\begin{equation}\label{primo}
 \left \langle  r,\theta,\phi\left  | J-\frac{1}{2},J_z -\frac{1}{2}  \right\rangle \right . =f_1^\mp(r)
\begin{pmatrix}
0\\
-\cos\theta\sin^{n-1}\theta e^{i(n-1)\phi}\\
0\\
\sin^n\theta e^{in\phi}
\end{pmatrix}
+
f_2^\mp(r)
\begin{pmatrix}
\sin^{n-1}\theta e^{i(n-1)\phi}\\
0\\
0\\
0
\end{pmatrix}\:.
\end{equation}
\end{widetext}
Here  $f^-(r) $ and $ f^+(r) $ are radial functions  localized at the boundary  for $r<R$ and  $r>R$, respectively, and they satisfy the equations:
\begin{equation}\label{sys}
\begin{split}
(E\mp\Delta)f_2^\mp&=-i\partial_r f_1^\mp-\frac{i}{r}(n+1)f_1^\mp\:,\\
(E\pm\Delta)f_1^\mp&=-i\partial_r f_2^\mp+\frac{i}{r}(n-1)f_2^\mp\:.
\end{split}
\end{equation}
 The system can be decoupled in a pair of  Bessel equations
\begin{equation}
\begin{split}
\frac{d^2}{dr^2}f_1^\pm+\frac{2}{r}\frac{d}{dr}f_1^\pm-\left[(\Delta^2-E^2)+\frac{n(n+1)}{r^2}\right]f_1^\pm&=0\:,\\
\frac{d^2}{dr^2}f_2^\pm+\frac{2}{r}\frac{d}{dr}f_2^\pm-\left[(\Delta^2-E^2)+\frac{n(n-1)}{r^2}\right]f_2^\pm&=0\:,
\end{split}
\end{equation}
whose solutions are:
\begin{equation}\label{sol}
\begin{split}
& if\; r<R \:\: \:\:\: \:\:\: \:\:\: \:
\begin{cases}
f_1^-(r)=-i  \frac{\Delta-E}{\kappa}\: C^-\:  i_{n}(\kappa r)\\
f_2^-(r)=C^{-}\:  i_{n-1}(\kappa r)
\end{cases}\\
&if\; r>R\:\: \:\:\: \:\:\: \:\:\: \:
\begin{cases}
f_1^+(r)=- i C^+  \: \frac{\Delta+E}{\kappa}k_{n}(\kappa r)\\
f_2^+(r)=C^{+} \: k_{n-1}(\kappa r)
\end{cases}
\end{split}
\end{equation}
where $i_n\:,k_n$ are the modified spherical Bessel functions:
\beq
i_n(x) \equiv \sqrt{\frac{\pi}{2x}}I_{n+\frac{1}{2}}(x)\:\: , \:\:\:  \:\:\: \:\:\: \:\:\:
k_n(x)\equiv \sqrt{\frac{\pi}{2x}}K_{n+\frac{1}{2}}(x) \: .
\end{equation}
The  matching conditions can be written written using Eq. \eqref{sol}
\begin{equation}
\begin{cases}
-iC^-\: (\Delta-E) \: i_{n}(\kappa R)=- iC^+ \: (\Delta+E) \: k_{n}(\kappa R)\:,\\
C^- \: i_{n-1}(\kappa R)=C^+ \: k_{n-1}(\kappa R)\:,
\end{cases}
\end{equation}
which give rise to an implicit equation for the eigenenergies of the system:
\begin{equation}
\frac{\Delta -E}{\Delta +E }  =- \frac{k_{n}(\kappa R)\: i_{n-1}(\kappa R)}{ i_{n}(\kappa R)\: k_{n-1}(\kappa R)}\:.
\label{det}
\end{equation}
This equation, in the limit $\Delta R\rightarrow\infty$,  gives the admissible values of the energy:
 \beq
 E_n = \pm  n \frac{\hbar v_F}{R}  \:\:\: ,  \:\:\: \:\:\: \:\:\:  \:\:\: n =1 , \: ...  \: , n _{max}
 \eneq
 which is reported in Sec. II.
\\
\\
\section{ Derivation of the elastic $t-$matrix for   scattering off  a  gaussian bump}
 We derive the elastic $t-$matrix when a  localized  deformation is present on the surface of  a TI. As in
\cite{JCV07} we consider a two-dimensional spatial sheet modeled on a
two-dimensional axial symmetric manifold with a single gaussian
bump.  The axial symmetric gaussian surface
may be represented in Minkowski space-time by the function
$\phi=\left(t,x,y,h(r)\right)$ with $r^2= x^2+y^2$. From  Eqs. \eqn{metric}, we may read the tetrads
\begin{equation}
\begin{split}
{e_x}^1 &=   \frac{\cos\theta}{\sqrt{1+f(r)}},\,\:\: {e_y}^1 =\frac{\sin\theta}{\sqrt{1+f(r)}} \:\: ,\\
{e_x}^2 &= -\frac{\sin\theta}{r}  \: ,\:\qquad{e_y}^2 = \frac{\cos\theta}{r} \:\: .
\label{eq:vie}
\end{split}
\end{equation}
The Dirac equation on a radially symmetric manifld is
\begin{widetext}
\bea
 -i\left[\frac{\sigma ^r}{\sqrt{1+f(r)}}\del_r +
\sigma^\theta\left (\frac{1}{r}\del_\theta+\frac{i}{2r}\left(
1-\frac{1}{\sqrt{1+f(r)}}\right)\sigma^z \right) \right]\Psi=E\Psi\:\: ,
\label{stadi}
\enea
\end{widetext}
The gauge potential in \eqref{stadi} is the spin connection
 \bea
 \Gamma_\mu= \frac{i}{2}\left( 1-\frac{1}{\sqrt{1+f(r)}}\right)\sigma^z \delta_{\mu 2}.\label{Gamma}
 \enea
We pose $ \Psi=\Phi  \:  \exp { \int_r^{+\infty} dr'\: A_\theta (r' ) } $ with $A_\theta $ is the spin connection above.   The  $m$  component of the spinor $\Phi $ has the form ($m$ is the angular momentum   integer):
\begin{equation}
\label{eq:ansatz1}
\Phi _m  ( r,\theta |   \vec{k} ,s) =
\left ( \begin{array}{c}
u_{sm}(r)\\
is \: v_{sm}(r) \: e^{i\theta}
\end{array} \right )
e^{im(\theta -\theta _k)}
\end{equation}
where $\theta _k $ is the angle that the direction of the $\vec{k} $ vector  of  the incoming wave forms with the polar axis.
Substituting  \eqref{eq:ansatz1} in the  Dirac  eq.\eqref{scur} and dropping the labels $s m$, we find that the functions $u(r),v(r) $ have to satisfy the following equations:
\begin{widetext}
\bea
\frac{1}{\sqrt{1+f}}\frac{d^2u(r)}{dr^2}+\frac{1}{r}\frac{du(r)}{dr}+\left( \frac{d}{dr}\frac{1}{\sqrt{1+f}}\right) \frac{du(r)}{dr}-\frac{m^2}{r^2}\sqrt{1+f} \: u(r)+k^2u(r)=0, \nn\\
\frac{1}{\sqrt{1+f}}\frac{d^2v(r)}{dr^2}+\frac{1}{r}\frac{dv(r)}{dr}+\left( \frac{d}{dr}\frac{1}{\sqrt{1+f}}\right) \frac{dv(r)}{dr}-\frac{(m+1)^2}{r^2}\sqrt{1+f}\:  v(r)+k^2 v(r)=0.
\label{sist}
\enea
\end{widetext}
Due to the symmetry of the problem is suitable to expand the Green's function for the flat space-time problem in polar coordinates
  \beq
 G(z,z') = \frac{1}{2\pi }\sum _{m=-\infty}^{+\infty}  \: e^{im (\theta -\theta' )} g_m(r,r' ).
 \eneq
 The Green function displaying  the correct jump of the derivative at $r=r'$ is:
\beq
g_m(x,x' ) = 2 \pi ^2  \: J_m \left ( x_< \right ) Y_m \left ( x_> \right ) \:\: .
 \eneq
 Here $ r_<  ( r_>) $ is the smaller (larger) of the two arguments $r,r'$. We now  specialize  the shape of the bump $h(r) $  to be the gaussian bump  $z(|\vec{r}|)$ defined in sec. III.A. This implies that $ f(r) = (4h^2r^2/ \ell ^4 ) e^{-2(r/\ell)^2 } $. We assume that the ratio $h/\ell$ is small, so that we can expand Eq.s \eqref{sist} by retaining just the lowest power of  $h/\ell $.
By comparison with the system for the flat space (i.e. $ f(r) =0 $),  we define  the perturbative potential :
  \beq
\frac{h^2}{\ell^2}V_m(r)  = \frac{2h^2}{\ell^4} r^2\: e^{-2r^2/\ell^2} \: \left [ \frac{d^2}{dr^2}  +\left (\frac{4}{r} -\frac{8r}{\ell ^2} \right ) \frac{d}{dr}
 -\frac{m^2}{r^2} \right  ] \:\: .
\eneq
In the Born approximation, the Dyson equation for $ e.g. \:\: u _{km} $ reads:
\begin{widetext}
\bea
u_{km}(r)=  J_m (kr) + \frac{h^2}{\ell^2} \int^{\infty}_{0} dr'\: r'  \: g_m (r,r' ) V_m(r') \: J_m (kr') \nn\\
=  J_m (kr) +   \frac{2h^2}{\ell^4} \:  \int^{\infty}_{0} dr' \:  g_m (r,r' )\:  {r'}^3 \: e^{-2{r'}^2/\ell ^2} \:  \left \{ k^2 + \left (\frac{3}{r'} -\frac{8r'}{\ell^2} \right ) \frac{d}{dr'}
\right \} \: J_m (kr')
\label{born1}
\enea
\end{widetext}
We have used the fact that  $J_m $ solves the Bessel differential equation to simplify the action of  $V_m $ on
$J_m$ itself.  By defining :
\begin{widetext}
\bea
 \tan \delta _m =    \frac{4\pi^2k^2h^2}{\ell^4}\:  \int^{\infty}_{0} dr'  \: J_m (kr' ) \:  {r'}^3\: e^{-2{r'}^2/\ell^2}\:  \left \{ 1 + \frac{1}{k^2} \left (\frac{3}{r'} -\frac{8r'}{\ell^2} \right ) \frac{d}{dr'}
\right \} \:  J_m (kr')  \:\: ,
\label{psh}
 \enea
 \end{widetext}
 the scattering state for  $r/\ell \to \infty $ takes the form  $
  u_{km}(r)\sim   J_m (kr) +    \tan \delta _m \:  Y_m (kr)  $. By exploiting the symmetry of the Dirac massless  equation with respect to replacements $ u\leftrightarrow v$, $m \leftrightarrow -m-1 $, it is easy to see, that the sums which include  $\delta _m $, for all $m$, are equal. Therefore, our result is valid  for both components of the spinor solution given by  Eq.(\ref{eq:ansatz}).

 The integrals of Eq.\eqref{psh}  can be evaluated analytically.  The asymptotic expansion of  the Bessel functions implies that,  far from the bump ( $r/\ell \to \infty $ ), the outgoing wave takes the form:
 \begin{widetext}
 \beq
 u_{km}(r) \sim _{r/\ell \to \infty}    \frac{1}{\sqrt{1+  \tan^2 \delta _m} } \: \cos \chi_m +   \frac{ \tan \delta  _{m}}{\sqrt{1+  \tan^2 \delta _m} } \: \sin \chi_m  \nn\\
  \equiv  \cos  \left ( \chi _m +  \delta _{m} \right )  \: \hspace{1cm}  \chi _m=  kr- \frac{m\pi}{2 } -\frac{\pi}{4} \:\: .
  \label{asi}
 \eneq
 \end{widetext}
We now evaluate the $t-$ matrix element for  a scattering event, in which  an incoming wave with wavevector $\vec{k} $ is scattered elastically by the gaussian bump and  a plane wave of wavevector $\vec{p}$ emerges. The $t-$matrix element is:
 \begin{widetext}
 \bea
 \langle p | t (\vec{k})|k \rangle  =   \left [ 1 + e^{-i(\theta _{p}- \theta _k )} \right ] \: \sqrt{ \frac{2 }{ \pi k R^2} }
  \: e^{i \pi /4}   \: \frac{1}{R}   \int_0^R  dr \:  e^{\frac{h^2}{ \ell ^4 } \int _r^{R} dr' \: {r' }^2\:  e^{-2(r' /\ell)^2 }}  \:\sum_{\substack{m}} [e^{2i\delta_m}-1] \: e^{-im (\theta _{p}-\theta _{k})  }\:
 \: .
  \nonumber
  \enea
  \end{widetext}
The space integral arises from the exponential  prefactor of $\Psi$ defined before Eq.\eqref{scur}.  To evaluate the  relaxation time formula  of Eq.(\ref{reltB}) we first perform the integral over the angle $\theta _p $ of  the square modulus of the angle dependent exponentials in the sums.   The integral is non vanishing only for $ m-m' = \pm 2 ,0 $.  By rearranging the sums then  eq. \eqref{reltB} is obtained.
\section{Semiclassical approximation}
 In this Section we present a tasteful derivation of  the classical  high energy limit for the relaxation time. The latter  can be obtained by assuming classical diffusion along the geodesic trajectories across the bump  and yields\cite{DHAB10}:
 \begin{equation}\label{stime}
\frac{1}{\tau}\approx  \frac{ v_F}{2A} \int db\,\theta^2(b) \: .
\end{equation}
Here $ b$ is the impact parameter of the incoming particle, while $\theta$ is the scattering angle and $A$ is the area of the sample.   The starting point is the usual expression for the relaxation time of  Eq.\eqref{reltB}:
\begin{equation}\label{time2}
\frac{1}{\tau} =  \frac{v_F}{k A}\int _0^\pi d\theta\: (1-\cos\theta)\left|\sum_m f_m(\theta)\right|^2
\end{equation}
 given in terms of the scattering amplitudes $f_m(\theta) = \left[e^{i2\delta_m}-1\right]e^{im\theta}$. At high energy, many $m$ terms contribute to the sum, so that we  take its continuum limit, which amounts to integrate over continuous values of the classical angular momentum  $m  = k b$. As forward scattering is excluded form Eq.\eqref{time2},  it is enough to   apply the saddle point approximation to the resulting  integral\cite{landau}:
\beq
\sum_m e^{i(2\delta_m+m\theta)} \approx  e^{i(2\delta_{m_0} + m_0\theta)}\times  \int  dm \: e^{i\left  . \frac{d^2\delta }{dm^2} \right |_{0}(m-m_0)^2}.
\label{gau}
\eneq
Here $m_0$ is the stationary point, which  solves  the saddle point equation:
\begin{equation}\label{sclass}
\left  . \frac{d\delta_m}{dm} \right |_{m_0}  -\frac{\theta}{2}=0.
\end{equation}
Derivation of  this equation, once more, provides a relation between the second derivative of the phase shift $\delta_m$ and the the angle $\theta$
\begin{equation}\label{cond}
\frac{d^2\delta_m}{dm^2} - \frac{1}{2}\frac{d\theta}{dm}=0.
\end{equation}
An analytical continuation in the complex $m$ plane allows us to make the integral converge.  Gaussian integration in Eq.\eqref{gau}  implies that
\begin{equation}
\left |\sum_m f_m(\theta) \right |^2 \approx \pi \: \left|\frac{dm}{d\theta}\right | \: ,
\end{equation}
thus yielding the expected result:
\begin{eqnarray}
\frac{1}{\tau} \sim \frac{v_F}{k}  \frac{1 }{A} \:\int d\theta(1-\cos\theta) \: \left |\frac{dm}{d\theta} \right | \nonumber\\
\sim \frac{v_F}{k}  \frac{1 }{A} \:\int dm\: \frac{\theta^2}{2}  \sim \frac{ v_F }{A} \:\int db\: \theta^2 (b)\: .
\end{eqnarray}
In the last equality the conservation of the angular momentum $m = k b $ has been exploited, together with    the remark that the scattering angle only depends on $b$ in the classical diffusion.  This reproduces the  desired high energy behavior.

The analysis of Eq.\eqref{reltB} provides a similar conclusion. The quantity $\delta_{m+1}-\delta_{m-1}$, appearing as the argument of the cosine, is $\approx 2\; d\delta_m /dm $.  At large incoming energies, $\delta_m \approx ( 2m +1 ) \pi/2$, which is consistent with the asymptotic form of the wavefunction given in Eq.\eqref{asi}.
To  lowest order  we get, according to Eq.\eqref{sclass},
\begin{equation}
\sin^2\delta_m-\cos{(\delta_{m+1}-\delta_{m-1})}\sin\delta_{m+1}\sin\delta_{m-1}\approx \frac{\theta^2}{2}\: ,
\end{equation}
so that we recover again
\begin{equation}
\frac{1}{\tau}\approx  \frac{v_F}{kA} \int dm\: \frac{\theta^2}{2}  \approx \frac{v_F}{A}\int db\; \theta(b)^2 \: .
\end{equation}


\begin{thebibliography}{21}
\bibitem[{ful()}]{fullerenes}
\bibinfo{note}{The carbon fullerenes admit a continuum description in terms of
  two Dirac equations on the surface of a sphere\cite{GGV92}. Lattice effects,
  such as scattering between the two equations, lead to corrections which scale
  as $(a / R )^2$, where $a$ is the lattice spacing, see\cite{GGV92}. We do not
  find terms of similar magnitude at the surface of a topological insulator.}
\end{thebibliography}

\end{document}